\documentclass[twocolumn]{aastex7}
\pdfoutput=1 
\usepackage[T1]{fontenc}
\usepackage[utf8]{inputenc}
\usepackage[english]{babel}
\usepackage{newtxtext, newtxmath}
\usepackage{enumitem}
\usepackage[figure,figure*]{hypcap}
\graphicspath{{figures/}}

\newcommand*{\kms}{\text{km}\,\text{s}\ensuremath{^{-1}}}
\newcommand*{\msun}{\ensuremath{M_{\odot}}}
\newcommand*{\perh}{\ensuremath{h^{-1}}}
\newcommand*{\mvir}{\ensuremath{M_\text{vir}}}
\newcommand*{\rvir}{\ensuremath{R_\text{vir}}}
\newcommand*{\vmax}{\ensuremath{V_\text{max}}}
\newcommand*{\mpeak}{\ensuremath{M_\text{peak}}}
\newcommand*{\ahalf}{\ensuremath{a_{1/2}}}
\newcommand*{\ampeak}{\ensuremath{a_{\mpeak}}}
\newcommand*{\vmaxmpeak}{\ensuremath{\vmax(\ampeak)}}
\newcommand*{\https}[1]{\href{https://#1}{\nolinkurl{#1}}}

\begin{document}

\renewcommand*{\sectionautorefname}{Section} 
\renewcommand*{\subsectionautorefname}{Section} 
\renewcommand*{\subsubsectionautorefname}{Section} 

\title{Correlation between Outlying Halo Abundance and Host Halo Properties}

\author[orcid=0009-0009-8272-5315]{Emily Sageser}
\affiliation{Department of Physics and Astronomy, University of Utah, Salt Lake City, UT 84112, USA}
\email{emily.j.sageser@gmail.com}

\author[orcid=0000-0002-1200-0820]{Yao-Yuan Mao}
\affiliation{Department of Physics and Astronomy, University of Utah, Salt Lake City, UT 84112, USA}
\email{yymao@astro.utah.edu}

\author[orcid=0000-0002-9820-1219]{Ekta Patel}
\altaffiliation{NASA Hubble Fellow}
\affiliation{Department of Physics and Astronomy, University of Utah, Salt Lake City, UT 84112, USA}
\email{ekta.patel@utah.edu}

\begin{abstract}
The correlations between dark matter halo properties and subhalo abundance, or occupation, have been studied extensively; however, existing studies have mainly focused on subhalos within the virial radius of the host halo. In this work, we quantify the correlation between host halo properties and the abundance of neighboring halos that reside right outside of the virial radius of the host halos. We compute the correlations between four host halo properties (half-mass scale, concentration, peak-mass scale, and spin) and the outlying halo occupation out to 1.5\,Mpc for Milky Way-mass host halos, and study how the correlation strength varies with radius.  We also investigate if the outlying halo occupation can provide information about the host halo properties. We find that host halo properties impact the neighboring halo abundance beyond the virial radius, and the locations at which the correlation peaks do not typically align with the virial radius or splashback radius. The behavior of this observed correlation as a function of radius, especially in the outskirts, is connected to the effect of halo assembly bias. However, there is no universal behavior when considering different host halo properties. Our results are the first to quantify the occupation variation of outlying halos beyond the virial radius. They provide the theoretical background for interpreting the observed satellite systems when the observed satellites are not strictly defined to be within the virial radius.
\end{abstract}


\section{Introduction}
\label{sec:intro}

In modern cosmology, dark matter forms a hierarchical structure where dark matter halos, dense regions of dark matter, are created through the gravitational interactions among dark matter particles \citep[see e.g.,][for a review]{2012AnP...524..507F}. Galaxies are believed to reside in these dark matter halos, and the connection between galaxies and the underlying halos, commonly called the galaxy--halo connection, has been the focus of a considerable amount of cosmological research \citep[e.g.,][]{2018ARA&A..56..435W}. 
Satellite galaxies, which are galaxies under the gravitational influence of larger nearby systems, are particularly interesting as they serve as the visible counterpart of the invisible dark matter hierarchical structure. As these satellite galaxies reside in dark matter subhalos, the studies around satellite galaxies can potentially help reveal the nature of dark matter \citep[e.g.,][]{2017ARA&A..55..343B,2021PhRvL.126i1101N}.

Cosmological simulations have shown that the number of subhalos, sometimes called subhalo abundance or occupation\footnote{In this work, we use these two words (``abundance'' and ``occupation'') interchangeably. They both refer to the number of (sub)halos.}, is highly dependent on the properties of the host halo. To the first order, the subhalo occupation is almost linearly proportional to the host halo mass \citep[e.g.,][]{2004ApJ...609...35K}. This proportional relation forms the basis of the satellite component in the halo occupation distribution (HOD) model, which models the number of satellite galaxies as a power law of the host halo mass, with the power index typically close to one \citep[e.g.,][]{2005ApJ...630....1Z,2005ApJ...633..791Z}.
While the HOD model can reproduce the observed galaxy population and its clustering signals reasonably well, more recent studies have also shown that the subhalo occupation is dependent on halo properties other than halo mass. 
In particular, halo formation time and halo concentration are known to strongly correlate with subhalo occupation at a fixed host halo mass \citep[e.g.,][]{2005ApJ...624..505Z,2015ApJ...810...21M,2018MNRAS.480.3978A}. 

This dependence of subhalo occupation on host halo properties other than mass is often referred to as occupation variation. Since satellite galaxies are thought to live inside subhalos, the subhalo occupation variation should translate to satellite occupation variation, sometimes also known as satellite galaxy assembly bias \citep[e.g.,][]{2014MNRAS.443.3044Z, 2016MNRAS.460.2552H, 2022MNRAS.516.4003W}. However, obtaining observational confirmation of the occupation variation is challenging. \citet{2022MNRAS.516.4003W} found marginal evidence from galaxy clustering statistics. More recently, the Satellites Around Galactic Analogs \citep[SAGA;][]{2017ApJ...847....4G,2021ApJ...907...85M,2024arXiv240414498M} and the Exploration of Local VolumE Satellites \citep[ELVES;][]{2020ApJ...891..144C,2022ApJ...933...47C} surveys have been studying Milky Way analog systems to identify dwarf satellite galaxies out to the virial radius. Among observational results is the characterization of how the number of satellite galaxies correlates with various host galaxy properties, which provides an indirect evidence for the subhalo occupation variation. 

The aforementioned theoretical and observational studies have all focused on the subhalos (or satellite galaxies) within the virial radius of the host halo. The virial radius is commonly considered the boundary of a dark matter halo, within which the dark matter particles are virialized. Practically, the virial radius is usually defined by a specific value of spherical overdensity \citep[e.g.,][]{1998ApJ...495...80B}.  
However, since halos are not completely isolated and in full equilibrium, other radius definitions have been proposed to represent the halo boundary \citep[e.g.,][]{2013MNRAS.430..725V,2024arXiv240604054S}.  
In particular, the splashback radius is thought to be a more physical definition of halo radius, as it is defined as the location at which the dark matter particles reach their apocenters and ``splash back'' into the gravitational potential. 
The splashback radius is generally around 1--2 times the distance of the virial radius \citep{2014ApJ...789....1D,2017ApJ...843..140D}, suggesting the region immediately outside the virial radius is still under the gravitational influence of the host halo. 
Studies using splashback radii have shown that this radius can encompass twice the abundance of subhalos compared to the virial definition \citep{2021ApJ...909..112D}.
It is, hence, very interesting to ask whether the (sub)halo occupation outside of the virial radius also exhibits occupation variation.

In this work, we aim to study the abundance of halos that reside right outside of the virial radius of a host halo and its associated occupation variation with respect to the host halo properties. 
In particular, we seek to determine if the outlying halo abundance is still under the influence of the host halo. If the host halo can impact the outlying halo abundance, we will quantify how such influence varies with radius and determine the location at which the outlying halo abundance exhibits the strongest correlations with host halo properties.
In addition, we will explore whether the outlying halo abundance at different radii can provide a way to constrain host halo properties such as host halo mass or concentration. These halo properties are important for comparison with theoretical work yet are difficult to measure. The outlying halo abundance, on the other hand, can  be viewed as an  approximation of the outlying galaxy abundance, which is observable. 

This work presents a detailed study of the occupation variation of outlying halos for the first time. This work is particularly relevant to recent observational efforts on satellite systems, such as the ELVES and SAGA surveys. While both surveys mainly focus on satellites within virial radii, the satellite definition uses a fixed physical radii, rather than the actual virial radius. 
More importantly, many recent observational works include ``satellite'' galaxies outside the virial radius \citep[e.g.,][]{2020A&A...643A.124K, 2022A&A...657A..54B, 2025PASA...42...26K, 2025A&A...693A..44M, 2025A&A...695A.106T}, which would reside in the outlying halos that we study in this work. As such, the results of this work provide a theoretical background for interpreting these observational studies. 

This work is organized as follows. \autoref{sec:methods} describes the simulation we use and our methodology. \autoref{sec:results} presents our main results, including the (sub)halo radial profiles split by selected host halo properties, the correlation between subhalo/outlying halo abundance and host halo properties, the radii where the strongest correlation occurs, and the attempt to use our findings to constrain halo properties. We discuss the implications of our results in \autoref{sec:discussion} and summarize in \autoref{sec:summary}.

\section{Methods}
\label{sec:methods}

\subsection{Simulations}
We use the Very Small MultiDark Planck Simulation (VSMDPL; \citealt{2012MNRAS.423.3018P, 2013AN....334..691R}) in this work. The VSMDPL simulation is a gravity-only $N$-body simulation that uses Planck Cosmology ($h=0.678$, $\Omega_\Lambda = 0.693$, $\Omega_m = 0.307$, $\Omega_b = 0.0482$, $n_s = 0.96$, and $\sigma_8 = 0.823$). It has a box size of 160\,Mpc\,\perh{} and a particle mass of $6.2 \times 10^6\,\msun\,\perh$. The simulation is processed through the Rockstar halo finder \citep{2013ApJ...762..109B} and Consistent Trees \citep{2013ApJ...763...18B}, and we use the present-day ($z=0$) halo catalog for this analysis. The virial radius, $\rvir$, is defined by a spherical overdensity $\Delta_c \approx 99$ of the critical density, which is the corresponding virial overdensity in this cosmology \citep{1998ApJ...495...80B}. 

\subsection{Terminology}
\label{sec:terms}
Throughout this article, we use the term ``host halo'' to refer to a dark matter halo that is the most massive halo within its radius, the term ``subhalo'' to refer to any smaller halo housed within the virial radius ($\Delta_c \approx 99$) of a more massive host halo, and the term ``outlying halo'' to refer to any smaller halo that resides outside of the virial radius of the host halo. 

From the halo catalog, we use the following halo properties: concentration ($c$), half-mass scale ($\ahalf$), peak-mass scale ($\ampeak$), and spin. These properties are defined as follows. 
\begin{itemize}[parsep=0pt]
    \item Half-mass Scale ($\ahalf$): The scale factor at which a halo first reaches half of its present-day mass.
    \item Concentration ($c$): The ratio of the virial radius to the scale radius of a dark matter halo, with the scale radius obtained by Rockstar via fitting the dark matter density profile to the Navarro--Frenk--White profile \citep{1996ApJ...462..563N}.
    \item Peak-mass Scale ($\ampeak$): The scale factor at which a halo has the maximal mass over its accretion history on the main branch.
    \item Spin: The dimensionless Peebles spin parameter \citep{1969ApJ...155..393P}, characterizing the angular momentum of the halo. 
\end{itemize}
All of these host halo properties are calculated with the virial radius definition. 

\subsection{Host Halo Selection}

In this work, we select isolated host halos that have $\mathtt{upid} = -1$ in the Rockstar and  is the more massive halo within 1.5\,Mpc. The isolation criterion ensures that none of the outlying halos will be more massive than their host halo. 
For most of the analyses, we constrain our host halo virial masses (\mvir{}) to be within $10^{12-12.05}$\,\msun\,\perh, which is a narrow range around the typical mass of a Milky Way analog, except for in Sections \ref{sec:radius-trend} and \ref{sec:probing} when we study the mass trend. 
In the VSMDPL simulation, we identify 1,392 isolated host halos in the Milky Way-mass range, and these halos have a median virial radius of 305.8\,kpc. 

\subsection{Subhalo and Outlying Halo Selection}

When selecting subhalos and outlying halos, we apply a threshold of $ \vmaxmpeak{} \geq 20$\,\kms. Here \vmaxmpeak{}  is defined as the maximum circular velocity at the time when the halo reaches its peak mass throughout its accretion history. We choose to apply the threshold on \vmaxmpeak{} because it is a proxy for luminosity thresholds \citep{2013ApJ...771...30R, 2016MNRAS.460.3100C, 2017ApJ...834...37L}. This threshold roughly corresponds to a stellar mass of 1--$4 \times 10^4$\,\msun, although the exact value heavily depends on the assumed faint-end slope of the stellar-to-halo mass relation. In the VSMDPL simulation, this $ \vmaxmpeak{} \geq 20\,\kms$ threshold results in 25.9 subhalos on average within the virial radius for our isolated Milky Way-mass hosts.
We have repeated our analyses with several different \vmaxmpeak{} thresholds. They all present the same qualitative trends and patterns, but higher \vmaxmpeak{} thresholds result in noisier results as one would expect. 

\section{Results}
\label{sec:results}

\subsection{Occupation Variation of Subhalos and Outlying Halos}
\label{sec:radprof}

\begin{figure*}[tb]
    \centering
    \includegraphics[width=\linewidth]{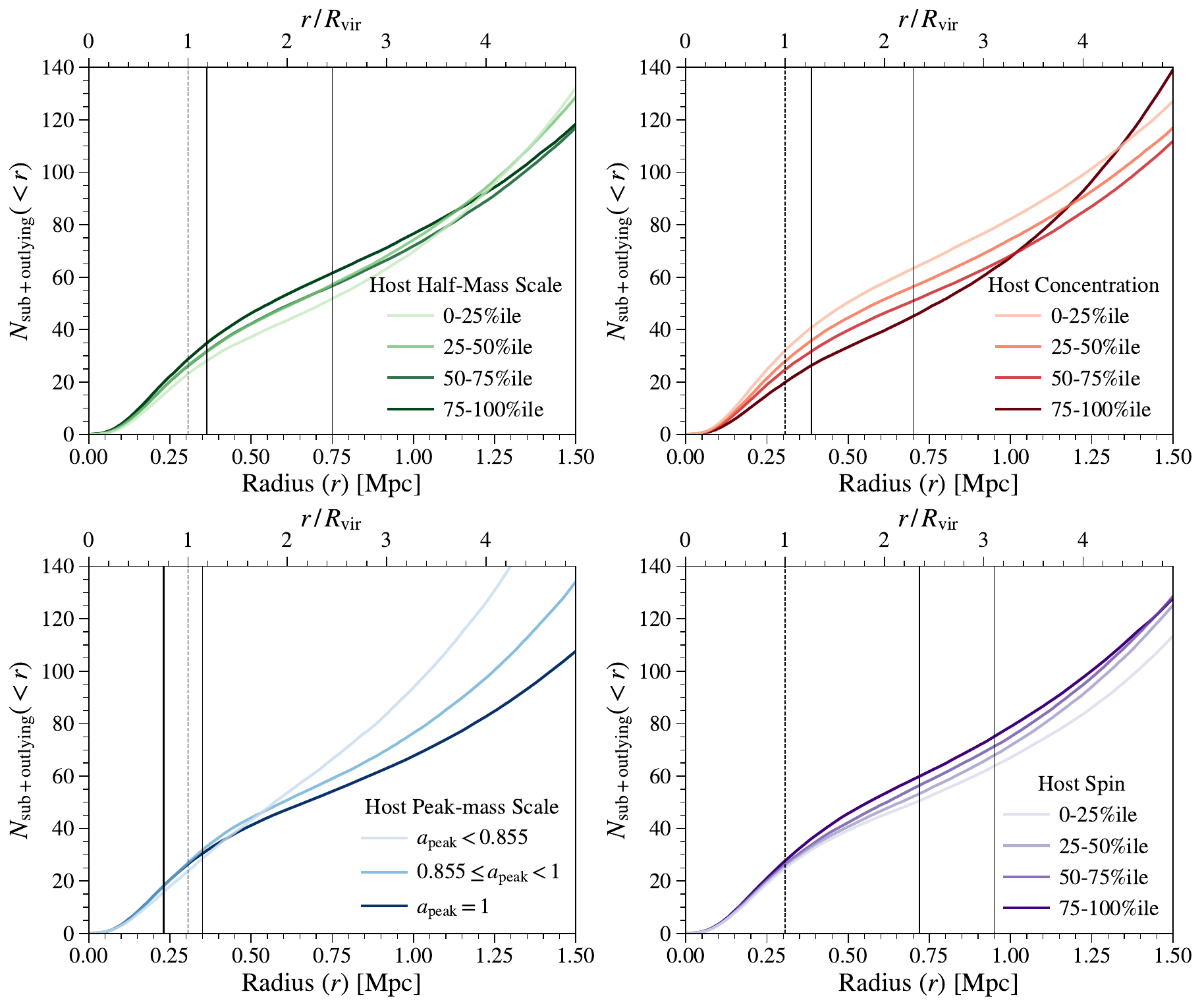}
    \caption{Average number of subhalos and outlying halos within radius $r$ (distance to the host halo center) split by four different host halo properties: half-mass scale (\textit{upper left; green}), concentration (\textit{upper right; red}), peak-mass scale (\textit{bottom left; blue}), and spin (\textit{bottom right; purple}), as defined in \autoref{sec:terms}. Curves of different shades on each plot represent the four quartiles, except for the case of peak-mass scale, where different shades correspond to the ranges specified in the corresponding legend. In each panel, the two solid vertical lines represent the region where the curves are roughly parallel to each other. The dashed vertical line represents the halo virial radius. The subhalos and outlying halos are selected with a \vmaxmpeak{} threshold of 20\,\kms. The host halo mass range is $10^{12-12.05}$\,\msun\,\perh, and they have a median \rvir{} of 0.3058\,Mpc. The upper $x$-axis shows the radius in multiples of the median \rvir{} value.}
    \label{fig:radial-dist}
\end{figure*}

The main goal of this work is to extend the study of subhalo occupation variation, the correlation between subhalo counts and host halo property, to outlying halos, which reside outside of the host halo's virial radius.
To this end, we plot the radial profiles, which show the cumulative number of subhalos and outlying halos as a function of radius (distance to the host halo center), in \autoref{fig:radial-dist}. The radial profiles are further split by four distinct host halo properties: half-mass scale, concentration, peak-mass scale, and spin. The dashed black line in each panel indicates the median virial radius of 0.3058\,Mpc for these host halos that are in the Milky Way-mass range (\mvir{} = $10^{12-12.05}$\,\msun\,\perh). Each panel shows the result of dividing the host sample into quartiles of a given host halo property, with the exception of peak-mass scale, which is divided into three approximately equal groups since most halos have a peak-mass scale close to one.

First, we focus on the behavior of the radial profiles within the virial radius. \autoref{fig:radial-dist} shows that as the radius increases, the radial profiles split by quartiles begin to diverge, implying the existence of the subhalo occupation variation. Within the virial radius, subhalo occupation variation exhibits most strongly for host concentration, but only moderately for the host half-mass scale and host peak-mass scale. The radial profile split by host spin shows almost no variation inside the virial radius. Taking the concentration panel (upper right) as an example, host halos with the highest concentration (darkest red line) have a subhalo radial profile that grows the slowest. This result confirms previous studies on subhalo occupation variation: host halos with higher concentration or lower half-mass scale have fewer subhalos within the virial radius \citep[e.g.,][]{2005ApJ...624..505Z,2015ApJ...810...21M,2018MNRAS.480.3978A}. On the other hand, there is no strong subhalo occupation variation with respect to the peak-mass scale or spin within the virial radius.  

The main motivation of this work, however, is to explore how the occupation variation changes for outlying halos that are beyond the virial radius. \autoref{fig:radial-dist} quantifies radial profiles to 1.5\,Mpc, nearly five times the virial radius. For half-mass scale, concentration, and host spin, the occupation variation (of outlying halos) continues to grow beyond the virial radius (i.e., the quartile lines continue to diverge). For each host halo property, this growth ceases for a specific range of radii where the variation remains constant (i.e., the quartile lines become parallel). For both concentration and half-mass scale, the growth stagnates around 0.36--0.39\,Mpc (1.2--$1.3\rvir$). For host spin, the occupation variation continues to grow until 0.72\,Mpc ($2.4\rvir$), where the growth stagnates. In each panel of \autoref{fig:radial-dist}, the region where the quartile lines grow at a steady rate is marked with solid vertical black lines.

Note that in the region where the quartile lines are parallel, there is no additional variation in the outlying halo occupation. In other words, if one only looks at the number of outlying halos within this region, hosts in different quartiles would have similar numbers.  The total number of subhalos and outlying halos, however, still differs among hosts in different quartiles, but this difference entirely comes from occupation variation in the inner region (within the leftmost black line). 

As we continue to move outward, a third region where the variation starts to shrink is reached. The shrinkage can be seen most clearly in the case of the half-mass scale, where beyond 0.75\,Mpc ($2.5\rvir$), the quartile lines begin to converge and even intersect. In the case of concentration, only the highest quartile exhibits significant shrinkage. This shrinking region roughly starts at 0.7\,Mpc ($2.3 \rvir$) for concentration. For spin, shrinkage begins at 0.95\,Mpc ($3.1 \rvir$), except for the lowest quartile. We also observe that the highest concentration quartile and the lowest half-mass scale quartile behave quite differently from other quartiles. We will discuss this further in \autoref{sec:assembly-bias}.

Radial profiles split by the peak-mass scale (bottom left panel of \autoref{fig:radial-dist}) behave quite differently from other halo properties. There is no strong trend in the occupation variation in the inner region below the virial radius. At about 0.23\,Mpc ($0.8\rvir$), the lines of thirds become parallel to one another, and around 0.35\,Mpc ($1.1\rvir$), the occupation variation begins to increase and continues to do so until 1.5\,Mpc. 

We have also checked that the trends in the radial profiles observed in \autoref{fig:radial-dist} persist when we use different \vmaxmpeak{} thresholds for subhalos and outlying halos and when we select different mass ranges for the host halos. The specific locations at which the variation behavior changes differ when we change the mass ranges of the host halos, which we will explore in \autoref{sec:correlation}. 
In summary, we conclude that the occupation variation does exist for outlying halos as it does for subhalos. In addition, the location where the strength of occupation variation starts to decrease does not align with the virial radius.

\subsection{Correlation Strength as a Function of Radius}
\label{sec:correlation}

\begin{figure*}[tb]
    \centering
    \includegraphics[width=\linewidth]{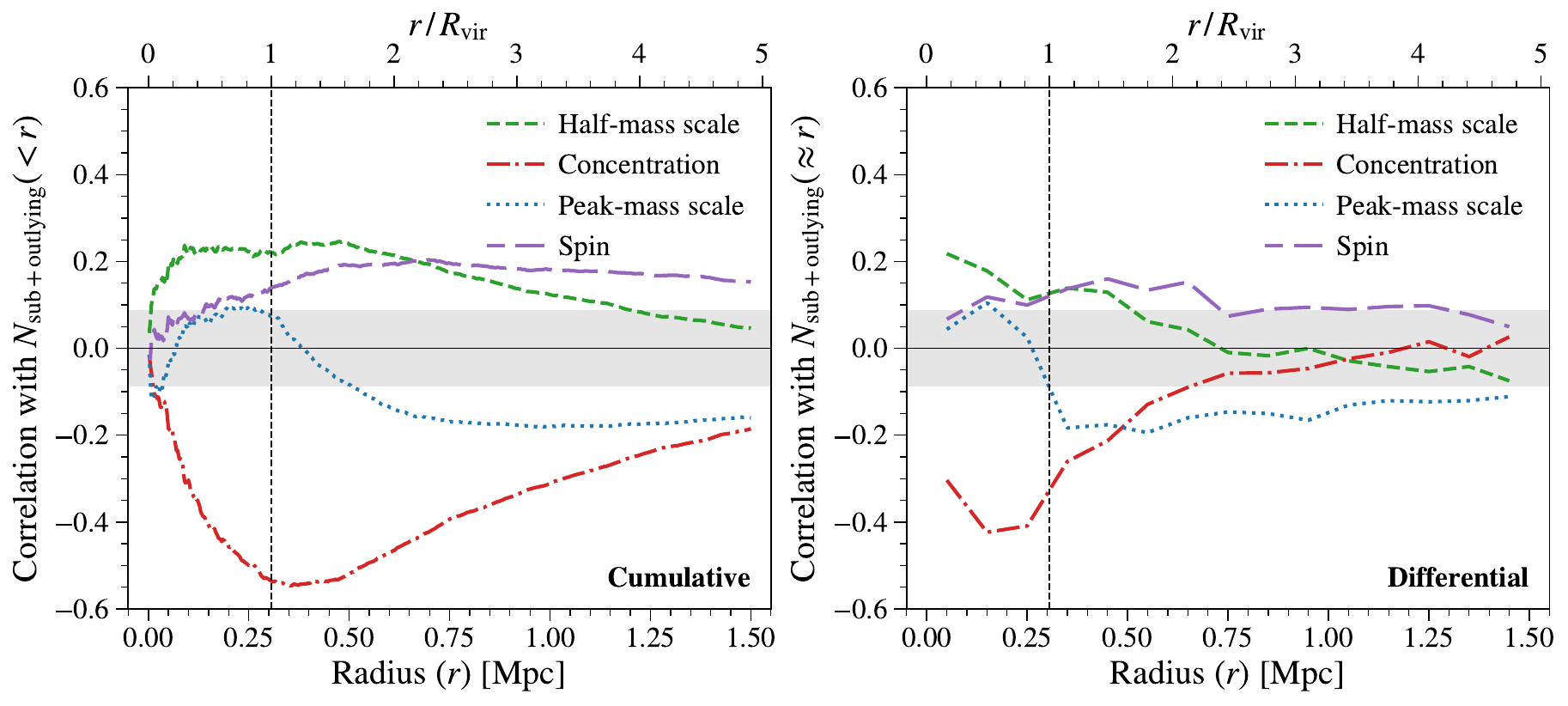}
    \caption{Spearman correlation between each of the four host halo properties and the number of subhalos and outlying halos within radius $r$ (cumulative; \textit{left}) and in radial bins (with bin size of 0.1\,Mpc; \textit{right}). The four host halo properties are half-mass scale (\textit{green dashed line}), concentration (\textit{red dash-dot line}), peak-mass scale (\textit{blue dotted line}), and spin (\textit{purple long dashed line}). The vertical black dashed line represents the halo virial radius. The horizontal gray band represents where the correlation has a $p$-value $< 0.001$. The subhalos and outlying halos are selected with a \vmaxmpeak{} threshold of 20\,\kms. The host halo mass range is $10^{12-12.05}$\,\msun\,\perh, and they have a median \rvir{} of 0.3058\,Mpc. The upper $x$-axis shows the radius in multiples of the median \rvir{} value.}
    \label{fig:correlation}
\end{figure*}

A different way to inspect the occupation variation is to calculate the correlation coefficient between the number of subhalos or outlying halos within certain radius and a host property at a fixed halo mass.
Specifically, for each value $r$ in the radius range of interest, we calculate the Spearman correlation coefficient between the number of subhalos and outlying halos within $r$ of the host and the chosen host property.
The number of subhalos and outlying halos is normalized by the host halo mass before the correlation coefficient calculation; however, whether the normalization is applied does not change the result due to the very narrow host halo mass range.
Note that in this study, all host halo properties are defined using \rvir{}, and their definitions do not change as we vary the radius for the correlation coefficient calculation. 
In this analysis, there are a total of 1,392 systems included, and a Spearman correlation coefficient of 0.0881 corresponds to a $p$-value of 0.001. 

The left-hand panel of \autoref{fig:correlation} shows how the correlation coefficient changes with radius. Here, a greater magnitude of the Spearman correlation implies that the number of subhalo and outlying halo varies more strongly with the host property in consideration. This behavior would manifest in \autoref{fig:radial-dist} as a wider spread between the quartile lines.\footnote{Note that we did not split the host peak-mass scales in quartiles in \autoref{fig:radial-dist} (because the majority of hosts have a peak-mass scale of 1); as such, the wider spread between the different peak-mass scale lines does not translate to stronger correlations.} We can see that the host concentration parameter shows the strongest negative correlation with the number of subhalos and outlying halos, and this negative correlation peaks around 0.36\,Mpc ($1.2\rvir$), the radius at which the lines in the upper left panel of \autoref{fig:radial-dist} become parallel with each other. In other words, this radius ($1.2\rvir$) is the boundary at which the concentration parameter has the strongest impact on the number of subhalos and outlying halos.

We also observe that, for each host property, the radius at which the  number of subhalos and outlying halos varies with the property the most is vastly different. For the half-mass scale, the maximal correlation occurs when the occupation is defined to be the number of subhalos and outlying halos within 0.36\,Mpc ($1.2\rvir$), which is a similar radius as the case of the concentration parameter. For the spin parameter, the correlation peaks at an even greater radius of 0.72\,Mpc ($2.4\rvir$). For the peak-mass scale, on the other hand, the maximal correlation occurs at a radius of 0.23\,Mpc ($0.8\rvir$). Notably, the radius at which the correlations peak for each host halo property corresponds to the radius at which the radial profiles in \autoref{fig:radial-dist} become parallel (left vertical solid line in each panel).

Thus far, we have only considered the cumulative definition of (sub)halo occupation (i.e., number of halos \textit{within} a certain radius). However, we can also consider the occupation in bins of radius to provide further insights into localized trends that we might not be able to observe in the cumulative case. The right-hand panel of \autoref{fig:correlation} shows the correlation between host properties and the number of subhalos or outlying halos within radial bins. There are 15 radial bins, and each bin has a width of 0.1\,Mpc. This bin size is chosen such that the correlation trend as a function of radius is not dominated by noise. 

With this differential view, we see that both the half-mass scale and concentration parameter have a strong impact on the number of subhalos in radial bins within \rvir. Their impact continues well beyond \rvir, until about 0.75\,Mpc ($2.5 \rvir$) where both correlations plateau. The strength of the correlations begins to decline around 0.4\,Mpc ($1.3 \rvir$) for half-mass scale and 0.15\,Mpc ($0.5 \rvir$) for concentration.  Similar to the left panel of \autoref{fig:correlation}, concentration is most strongly correlated with subhalo occupation within \rvir. The other two host properties behave quite differently. The host spin parameter shows a relatively constant but mild correlation with the number of subhalos or outlying halos at all radial bins.  The peak-mass scale exhibits a weak positive correlation with subhalo occupation within \rvir, but a moderate negative correlation with outlying halo occupation outside of \rvir. This last behavior is related to the effect of halo assembly bias, which we will discuss in detail in \autoref{sec:assembly-bias}.

\subsection{Radii where Host Properties Impact Occupation the Strongest}
\label{sec:radius-trend}

Figures \ref{fig:radial-dist} and \ref{fig:correlation} both demonstrate that outlying halo occupation outside of \rvir{} is indeed impacted by host properties. For convenience, we will call the radius within which the number of subhalos and outlying halos has the strongest correlation with a host property the ``strongest-correlation'' radius. This strongest-correlation radius is indicated as the leftmost black vertical line in each panel of \autoref{fig:radial-dist}, and as the location at which the correlation has the strongest magnitude in the left-hand panel of \autoref{fig:correlation}.

We observe that the strongest-correlation radius differs for different host properties, and does not coincide with \rvir{}. This raises the question of whether the splashback radius, which is usually considered a more physical definition of a halo boundary, may explain what we have observed. The splashback radius is the location at which the dark matter particles associated with a halo reach their apocenters. Studies have shown that the splashback radius of a host halo is approximately 1--2 times that of the virial radius \citep{2014ApJ...789....1D, 2017ApJ...843..140D}, and that host halos with greater mass have smaller ratios of splashback radii to the virial radius \citep{2014ApJ...789....1D, 2017ApJ...843..140D}. It would appear that the strongest-correlation radius may be connected with the splashback radius, given that our strongest-correlation radii are approximately 0.8--2.4\rvir.

\begin{figure}[tb]
    \centering
    \includegraphics[width=\linewidth]{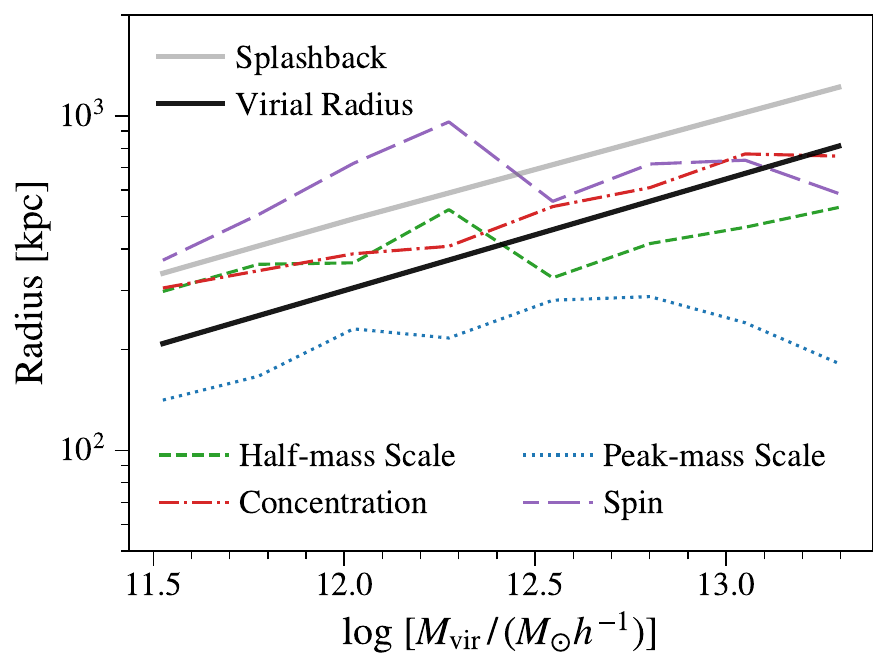}
    \caption{Relation between the strongest-correlation radius and host halo mass. The strongest-correlation radius is the radius within which the number of subhalos and outlying halos correlates most strongly with a halo property. The halo properties we analyzed are half-mass scale (\textit{green dashed line}), concentration (\textit{red dash-dot line}), peak-mass scale (\textit{blue dotted line}), and spin (\textit{purple long dashed line}). The virial radius (\textit{black solid line}) and the splashback radius (\citealt{2015ApJ...810...36M}; \textit{gray solid line}) are shown for comparison. Subhalos and outlying halos are selected by a \vmaxmpeak{} threshold of 20\,\kms. Host halo mass ranges are listed in \autoref{sec:radius-trend}. }
    \label{fig:radii}
\end{figure}

To test this idea, we repeat our analysis with a few different halo mass ranges to see how the strongest-correlation radius evolves with host halo mass and how it compares with the evolution of splashback radius. 
The mass ranges we analyze are $\log_{10} \left[ \mvir{} / (\msun\,\perh) \right] \in $ 
$[11.5, 11.55)$,
$[11.75, 11.8)$,
$[12, 12.05)$,
$[12.25, 12.35)$,
$[12.5, 12.6)$,
$[12.75, 12.85)$, 
$[13, 13.1)$, and 
$[13.25, 13.35)$. 
\autoref{fig:radii} shows how the strongest-correlation radius evolves with host halo virial mass, for each of the four host halo properties we have explored. The virial radius and splashback radius as functions of halo mass are also shown for comparison. The splashback radii shown in \autoref{fig:radii} are obtained from Figure~6 of \citet{2015ApJ...810...36M}.

As concluded from \autoref{fig:correlation}, the strongest-correlation radius is different for different host properties, but generally, it increases as the host halo mass increases. 
Some non-monotonic behaviors can be observed, but they are likely due to imperfect identification of the strongest-correlation radii. We applied some smoothing to \autoref{fig:correlation} to the identification of the strongest-correlation radii, but the shot noise (from halo counting) still impacts some mass bins significantly. In \autoref{fig:correlation-masses}, we include the full set of correlation--radius curves that are used to generate \autoref{fig:radii} for readers who wish to inspect further.

For host concentration and half-mass scale, the strongest-correlation radius is mostly between the virial and splashback radii. On the other hand, the host peak-mass scale positively correlates with the subhalo occupation only in inner regions (within \rvir{}), so the strongest-correlation radius for peak-mass scale is always smaller than the virial radius. 
The strongest-correlation radius for spin can go beyond the splashback radius, although the correlation strength is generally weak for host spin (recall from \autoref{fig:correlation}). 
In all four cases, the strongest-correlation radii do not fully match either the virial radius or the splashback radius. This is not entirely unexpected given the competing effect of the halo assembly bias and occupation variation. We will discuss further in Sections~\ref{sec:assembly-bias} and \ref{sec:splashback-radius}.

\subsection{Probing Host Halo Mass and Secondary Properties}
\label{sec:probing}

\begin{figure*}[tb]
    \centering
    \includegraphics[width=\linewidth]{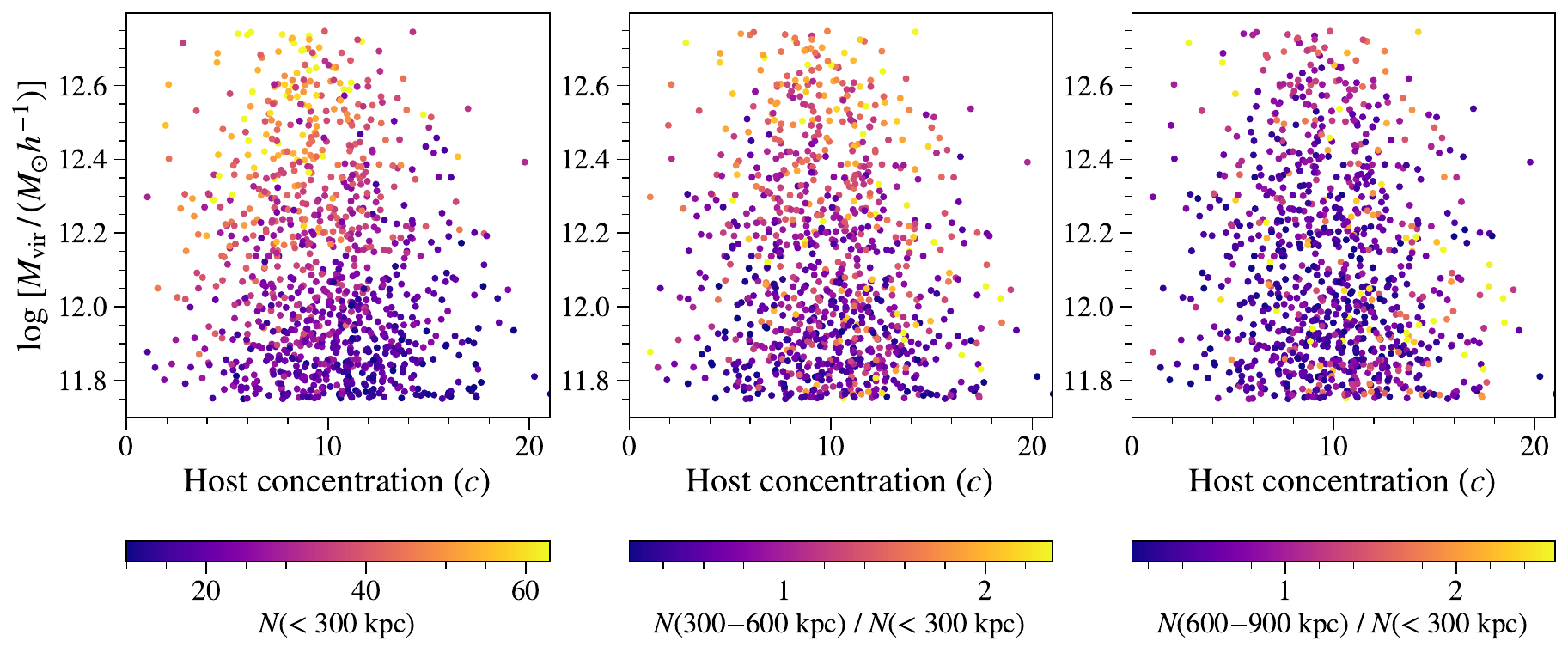}
    \caption{Joint distribution of host halo mass and concentration, with each host colored by different halo occupation statistics: the number of subhalos and outlying halos within 300 kpc (\textit{left}), number of subhalos and outlying halos between 300 and 600 kpc divided by the number of subhalos and outlying halos within 300 kpc (\textit{middle}), and the number of subhalos and outlying halos between 600 and 900 kpc divided by the number of subhalos and outlying halos within 300 kpc (\textit{right}). Subhalos and outlying halos are selected with a \vmaxmpeak{}  threshold of 20\,\kms. The gradient direction of the color demonstrates the dependence of the halo occupation statistics on the host halo mass and concentration.}
    \label{fig:mass-trend}
\end{figure*}

Observationally, halo occupation can often be directly measured via galaxy occupation, while halo mass and other halo properties are more difficult to measure. 
As such, it would be a powerful tool if one could use measurements of halo occupation to infer other host halo properties. 
\autoref{fig:correlation} demonstrates that the correlation between a given halo property and the number of subhalos and outlying halos within a given radius can have different strengths as the radius varies. This effect hints at the possibility that some combinations of the measurements of halo occupation in specific radial bins may be a good proxy for a single halo property. 

In this work, we conduct a preliminary exploration of this possibility. 
Specifically, we investigate whether we can find a specific combination of halo occupation at different radii such that the combination only correlates with halo mass or a secondary halo property (e.g., concentration). In this exploration, we choose several specific radius values, rather than multiples of the virial radius, to mimic observation of galaxy counts where the true virial radius of the host halo is unknown.

In the left-hand panel of \autoref{fig:mass-trend}, we see the typical joint dependence of the subhalo occupation on both host halo mass and concentration, indicated by the color gradient going diagonally from upper left to lower right. 
Here the halo mass range is $10^{11.75-12.75}$\,\msun\,\perh, and it roughly corresponds to the plausible halo mass range for the luminosity-selected Milky Way-mass galaxies in the SAGA Survey \citep{2024arXiv240414498M}. 
For the commonly used definition of the subhalo occupation within 300\,kpc (for Milky Way-mass systems; e.g., \citealt{2017ApJ...847....4G}), we cannot fully ascertain whether the occupation we observe comes from a higher-mass halo with a higher concentration or a lower-mass halo with a lower concentration. 
The goal of this exercise is to find a combination of subhalo and outlying halo counts in different radial bins such that this combination has little correlation with halo mass or concentration (i.e., color gradient goes vertically or horizontally). 

In the middle panel of \autoref{fig:mass-trend}, we calculate the ratio of the number of subhalos and outlying halos  between 300 and 600\,kpc to the number of subhalos and outlying halos within 300\,kpc. This particular combination removes most of the concentration dependence. While the resulting correlation with halo mass is slightly weakened, this combination seems to serve as a good proxy to estimate halo mass. 
In the right-hand panel of \autoref{fig:mass-trend}, we go further and calculate the number of subhalos and outlying halos between 600 and 900\,kpc to the number of subhalos and outlying halos within 300\,kpc.  In this case, we are able to eliminate most of the mass dependence; however, the remaining correlation between this ratio and the host concentration is rather weak. We only observe that hosts with high ratio values (points with the yellow color) tend to have higher concentration. 

In this preliminary exploration, we found that extending a satellite survey (such as SAGA or ELVES surveys) beyond the typical virial radius can potentially help obtain better constraints on halo mass or other halo properties. For example, based on the middle panel of \autoref{fig:mass-trend}, the ratio of galaxy counts between 300 and 600\,kpc to within 300\,kpc has little dependence on halo concentration, for halos in the $10^{11.75-12.75}$\,\msun\,\perh{} mass range (note that a $10^{12}\,\msun\,\perh$ halo has a $\rvir \approx 300$\,kpc).
We were not able to identify a combination that does not depend on halo mass but remains strongly correlated with halo concentration. 
We will discuss the observational implications of our findings in \autoref{sec:isolate-impact}, but will leave further investigation, including using machine learning methods to identify other promising combinations and testing against other host properties, to future work.

\section{Discussion}
\label{sec:discussion}

\subsection{Combined Effect of Halo Assembly Bias and Occupation Variation}
\label{sec:assembly-bias}

In this work, we mainly focus on halo occupation variation, which characterizes how the number of subhalos and outlying halos depends on a (non-mass) halo property at a fixed halo mass. A related but distinct effect, the halo assembly bias, characterizes how the matter density in the neighborhood of a halo depends on a (non-mass) halo property at a fixed halo mass \citep[e.g.,][]{2005MNRAS.363L..66G,2006ApJ...652...71W,0801.4826}. While these two effects are distinct in their definitions, the halo assembly bias effect can manifest as occupation variation when we consider the outlying halos. 

\autoref{fig:radial-dist} shows that, in the outermost region (greater than 1\,Mpc), systems with the highest concentration or lowest half-mass scale have a very different outlying halo distribution. These high-concentration (low half-mass scale) host halos have a larger number of outlying halos, in contrast to the fewer subhalos in the inner region.  This is a direct result of the combined effect of halo assembly bias and occupation variation. 

In the inner region, the effect of the subhalo occupation variation dominates. Late-forming or low-concentration halos tend to have more subhalos \citep[e.g.,][]{2005ApJ...624..505Z,2015ApJ...810...21M,2018MNRAS.480.3978A}. In the outer region, on the other hand, the effect of halo assembly bias is stronger than subhalo occupation variation. One way the halo assembly bias manifests is that early-forming (or high-concentration) halos cluster together more than late-forming (or low-concentration) halos do \citep{2005MNRAS.363L..66G, 2006ApJ...652...71W, 2008MNRAS.389.1419L}. This picture is consistent with our observation. 

Furthermore, \autoref{fig:radial-dist} shows two other intriguing features. First, the radius beyond which the halo assembly bias effect becomes more important than the subhalo occupation variation is much further out than the virial radius. In both the cases of concentration and half-mass scale, the inversion radius (e.g., the radius at which the highest concentration line crosses the other lines in the upper right panel of \autoref{fig:radial-dist}) is about 3--4 times that of the virial radius. Second, the halo assembly bias effect appears to only manifest in systems with the highest concentration or lowest half-mass scale. This feature is particularly apparent in the upper right panel of \autoref{fig:radial-dist}, where systems in the lower three quartiles of concentration remain in the same order of their outlying halo occupation.

The trade-off in subhalo occupation variation and halo assembly bias observed in our results is likely connected to the findings of \citet{2020MNRAS.493.4763M}, who showed that halo assembly bias is in part caused by splashback halos being identified as main halos (host halos). Even with the isolation criterion we imposed (no halos that are more massive within 1.5\,Mpc), it is possible that some of the main halos are still under the gravitational influence of a more massive system. 
\citet{2020MNRAS.493.4763M} shows that this situation is most likely when the main halos in consideration have a very high concentration (see their Figure 6). 

In fact, about 32.4\% of the hosts in our analysis (isolated, Milky Way-mass) have a peak-mass scale less than 1, which means they reach their peak mass before the present day. Among these hosts, the median peak-mass scale is $a=0.855$ ($z=0.17$). These hosts have a much higher number of outlying halos beyond 2\rvir{} (as seen in the lower left panel of \autoref{fig:radial-dist}). These systems with low peak-mass scales are the main contributors to the high-concentration systems that exhibit a similar radial distribution. As demonstrated in \autoref{fig:c_vs_ampeak}, high-concentration hosts are more likely to have a non-unify peak-mass scale, and when these hosts are removed from our sample, the high-concentration hosts' outlier behavior in the radial profile (as shown in \autoref{fig:radial-dist}) disappears. Our finding is consistent with the conclusion of \citet{2020MNRAS.493.4763M}. 

For both halo spin and peak-mass scale, there is no strong occupation variation within the virial radius. In other words, the host's spin and peak-mass scale impact the number of subhalos within the virial radius only very weakly. 
Outside of the virial radius, however, these two properties behave quite differently. 
For halo spin, we do not observe the ``inversion'' effect in the radial distribution shown in \autoref{fig:radial-dist}. 
\autoref{fig:correlation} further shows that the halo spin parameter correlates with the number of outlying halos with a roughly constant strength at all radii. This observation is consistent with our previous finding that the halo spin does not introduce strong halo assembly bias in the $10^{12}\,\msun\,\perh$ mass regime \citep{2017MNRAS.472.1088V}. 
Peak-mass scale, on the other hand, correlates negatively with the outlying halo occupation at large radii. This effect is likely caused by the crowded environment in which these low peak-mass scale systems reside, as discussed above. 

\subsection{Connection with the Splashback Radius}
\label{sec:splashback-radius}

\autoref{fig:correlation} shows that the peak correlation strength between host halo properties and number of subhalos and outlying halos does not necessarily occur at the virial radius of the host halo, and the radius where the peak correlation strength occurs, the strongest-correlation radius, differs across specific halo properties. \autoref{fig:radii} further demonstrates that the strongest-correlation radius does not fully agree with the splashback radius, but generally lies in between the virial radius and the splashback radius, with the exception of our results for the peak-mass scale. 

The fact that the strongest-correlation radii for some halo properties are outside of the virial radius implies that inner properties of a halo are still connected to the halo's environment. This implication is not surprising for several reasons. First, a halo and its environment are certainly linked via their neighboring initial conditions \citep[cf.][]{10.1142/S0218271807010511,2018PhDT.......204C,2020MNRAS.493.4763M}. Second, the gravitational potential of the main halo also extends beyond the virial radius and can impact outlying halos before they fall into the virial radius \citep{2014ApJ...787..156B}. As such, the virial overdensity, as a definition for halo radius, does not necessarily capture the physical processes that naturally set the halo boundary. This conclusion is consistent with previous studies on the splashback radius \citep{2021ApJ...909..112D}. 

The location of the strongest-correlation radius is a combined effect of both internal processes (occupation variation) and large-scale environment (assembly bias), as discussed in \autoref{sec:assembly-bias}. Hence, one may not expect the strongest-correlation radius to coincide completely with the splashback radius.
We will leave further exploration of the relation between the strongest-correlation radius and the splashback radius to future studies, but we provide some tentative explanations below. 

First, the splashback radius--mass relation shown in \autoref{fig:radii} is the average relation. Individual halos may have slightly different splashback radii that better match the strongest-correlation radii. Second, we use the original halo properties (i.e., with the virial definition) to calculate the correlation rather than fully adjust the radius definition for both halo properties and subhalo occupation. Third, subhalos experience dynamical friction, which depends on mass, and therefore, the response to the halo potential is slightly different for subhalos than for dark matter particles, which are used to define the splashback radii.

\subsection{Isolating the Impact of a Halo Property on Halo Occupation}
\label{sec:isolate-impact}

Directly measuring halo properties is a very challenging task. 
Even for basic properties such as the total halo mass and concentration parameter (derived from mass profile), we can only measure them for certain systems: very massive systems where gravitational lensing measurements are possible or very nearby systems where we have access to the velocity of stars, tidal debris, or satellite galaxies within the systems \citep[and references therein]{2016ARA&A..54..529B,2020SCPMA..6309801W}. 
If we would like to measure the halo mass or mass profile for a distant Milky Way-mass galaxy, usually, we can only obtain a ballpark value using the stellar mass--halo mass relation \citep[e.g.,][]{2013ApJ...770...57B} or brightest satellites \citep[e.g.,][]{2016ApJ...832...39L}.

Obtaining better estimates of various halo properties is very desirable as it can enable more detailed studies of the galaxy--halo connection.
One possible approach is to use the satellite count to estimate halo properties because the former is directly linked to subhalo occupation, which depends on host halo properties. 
However, subhalo occupation depends on multiple host halo properties simultaneously (as demonstrated in the left-hand panel of \autoref{fig:mass-trend}). 
Hence, our preliminary study in \autoref{sec:probing} aims to isolate the impact of specific host halo properties on subhalo occupation.

In the case of separating the effect of halo mass and halo concentration, we found that it is difficult to eliminate the mass dependence fully. While we found a combination of subhalo and outlying halo counts that can mostly eliminate the mass dependence, the said combination only weakly correlates with halo concentration. 
However, in this study, we have not tried a machine learning approach, which can provide a systematic search to find the best combination of subhalo occupation at different radii to eliminate mass dependence (or dependence on other halo properties). We will leave this to future work. 

If we can find some combination of halo occupation at different radii that correlates only with one specific halo property, this result will have significant observational implications. Since halo counts can be observed as galaxy counts, we can potentially measure halo properties for systems where we have robust galaxy counts that go beyond the host virial radius. These systems include the Milky Way-mass systems from the ELVES and SAGA surveys and the more recent observations that include galaxies outside the virial radius \citep[e.g.,][]{2020A&A...643A.124K, 2022A&A...657A..54B, 2025PASA...42...26K, 2025A&A...693A..44M, 2025A&A...695A.106T}.
If we repeat our analysis at the cluster mass scale, we may be able to better connect galaxy richness and halo mass for galaxy clusters.  
At the same time, this study also highlights the importance of observationally identifying galaxies outside of a host halo’s virial radius, as our results suggest those galaxies are still well connected with the host system and can, therefore, be used to constrain halo properties.

\section{Conclusions}
\label{sec:summary}

In this work, we study how the behavior of subhalo occupation variation (i.e., the dependence of the number of subhalos on halo properties other than halo mass) would change with different halo radius definitions. In particular, we studied the impact of four host halo properties: half-mass scale, concentration parameter, peak-mass scale, and spin parameter. Here, we summarize our main findings. 

\begin{itemize}[parsep=0pt]
    \item All four halo properties impact the number of outlying halos beyond the virial radius (\autoref{fig:radial-dist}). In other words, the halo occupation variation still exists when considering the outlying halos. 
    \item The radius at which the occupation variation of subhalos or outlying halos is strongest does not align with the virial radius (\autoref{fig:correlation}) or the splashback radius (\autoref{fig:radii}; see also \autoref{sec:splashback-radius}). However, in the case of host concentration, the strongest-correlation radius typically lies in between the virial radius or splashback radius  (\autoref{fig:radii}).  
    \item The relation between the strongest-correlation radius and host halo mass (see \autoref{fig:radii}) varies for the four different host halo properties we studied. There does not appear to be a universal behavior.  
    \item The ratio of the number of outlying halos between \rvir{} and 2\rvir{} to the number of subhalos within \rvir{} appears to be a better indicator of the host halo mass, as this ratio has a weaker dependence on halo concentration than the number of subhalos within \rvir{} has (\autoref{fig:mass-trend}).
    \item The behavior of the halo occupation variation as a function of radius, especially in the outskirts, is connected to the effect of halo assembly bias (\autoref{sec:assembly-bias}). 
\end{itemize}

While it is not surprising that outlying halos are still connected with the host halo's intrinsic properties, this work shows that it is unlikely to find a single halo radius definition that can simply explain the intricacy of the observed correlation. This result is similar to the finding of \cite{2017MNRAS.472.1088V}. 

Note that we choose to focus our study on the Milky Way halo mass scale, the same analysis can be applied to different mass scales, for example, for studying cluster membership \citep[cf.][]{2410.20205}, but we will leave that for future work.

This work also shows the preliminary exploration of eliminating the dependence of halo occupation on host halo mass. Implications of this exploration include providing better estimates of various host halo properties, which can enable more detailed studies of the galaxy-halo connection in future observational research.

\begin{acknowledgements}
The authors thank Benedikt Diemer, Philip Mansfield, and the anonymous referee for their helpful comments on this manuscript. 
E.S.\@ was supported in part by the Undergraduate Research Opportunity Program and the Physics \& Astronomy Summer Undergraduate Research Fellowship at the University of Utah. 
E.P.\@ is supported by NASA through Hubble Fellowship grant \# HST-HF2-51540.001-A, awarded by the Space Telescope Science Institute (STScI). STScI is operated by the Association of Universities for Research in Astronomy, Incorporated, under NASA contract NAS5-26555.

This work made use of the VSMDPL simulation. 
The authors gratefully acknowledge the Gauss Centre for Supercomputing e.V. (www.gauss-centre.eu) and the Partnership for Advanced Supercomputing in Europe (PRACE, www.prace-ri.eu) for funding the MultiDark simulation project by providing computing time on the GCS Supercomputer SuperMUC at Leibniz Supercomputing Centre (LRZ, www.lrz.de).
The MultiDark Planck simulation suite has been performed by Gustavo Yepes at the SuperMUC supercomputer at LRZ using time granted by PRACE and project time granted by LRZ. Processed data of these simulations were created by P. Behroozi, S. Gottlöber, A. Klypin, N. Libeskind, F. Prada, V. Turchaninov and G. Yepes.

The support and resources from the Center for High Performance Computing at the University of Utah are gratefully acknowledged.
This research has made use of NASA's Astrophysics Data System.
\end{acknowledgements}

\software{
Numpy \citep{2020NumPy-Array},
SciPy \citep{2020SciPy-NMeth},
Matplotlib \citep{matplotlib},
IPython \citep{ipython},
Jupyter \citep{jupyter},
pandas \citep{2022zndo...3509134T},
easyquery (\https{github.com/yymao/easyquery}), and
adstex (\https{github.com/yymao/adstex})
}

\bibliographystyle{aasjournalv7}
\bibliography{references}

\counterwithin{figure}{section}

\appendix

\section{Additional Analyses}
\label{sec:appendix}

\begin{figure*}[tb!]
\centering
\includegraphics[width=\linewidth]{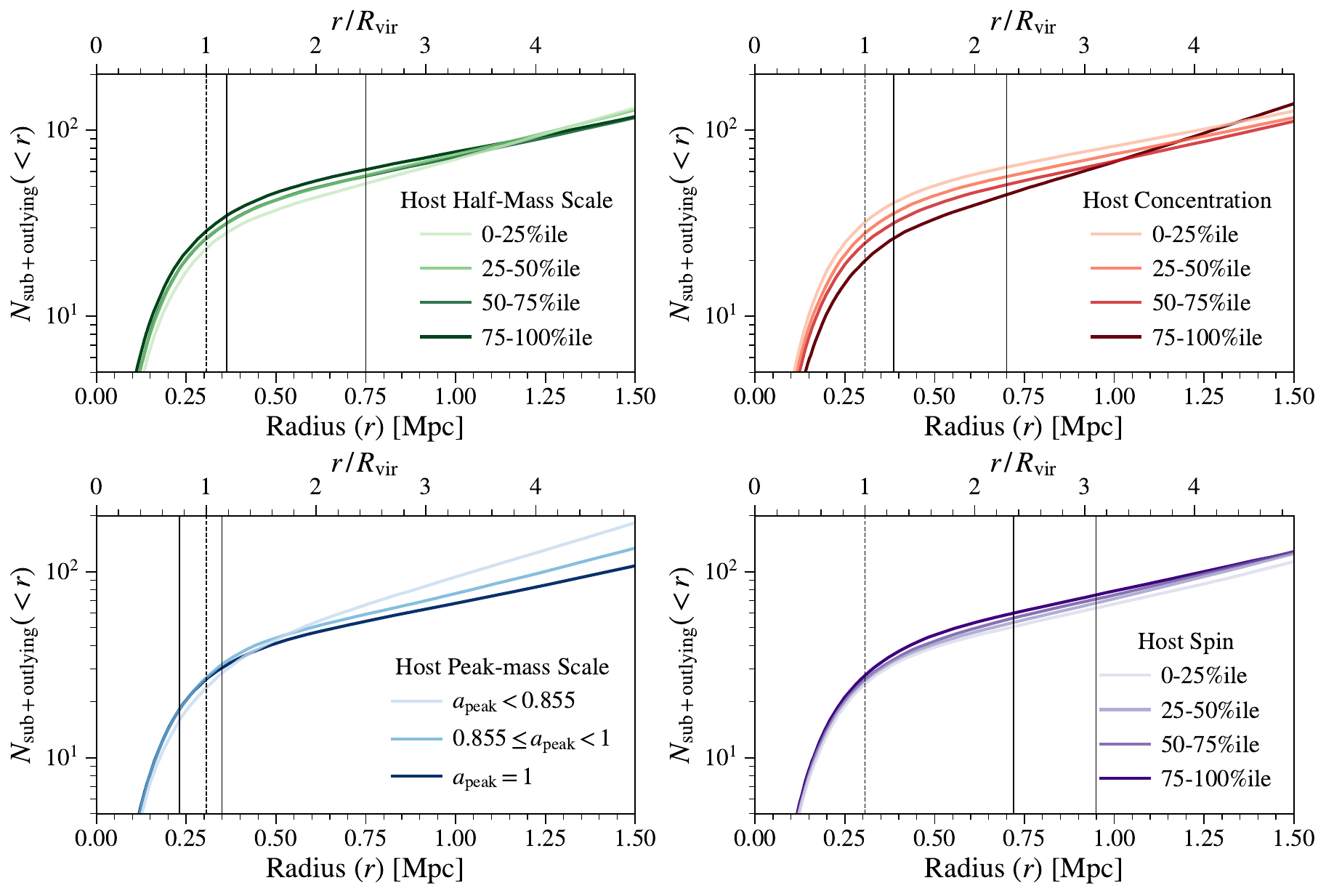}
\caption{Same as \autoref{fig:radial-dist} but with $y$-axis shown in the logarithmic scale.
\label{fig:radial-dist-log}
}
\end{figure*}

\begin{figure*}[tb!]
\centering
\includegraphics[width=\linewidth]{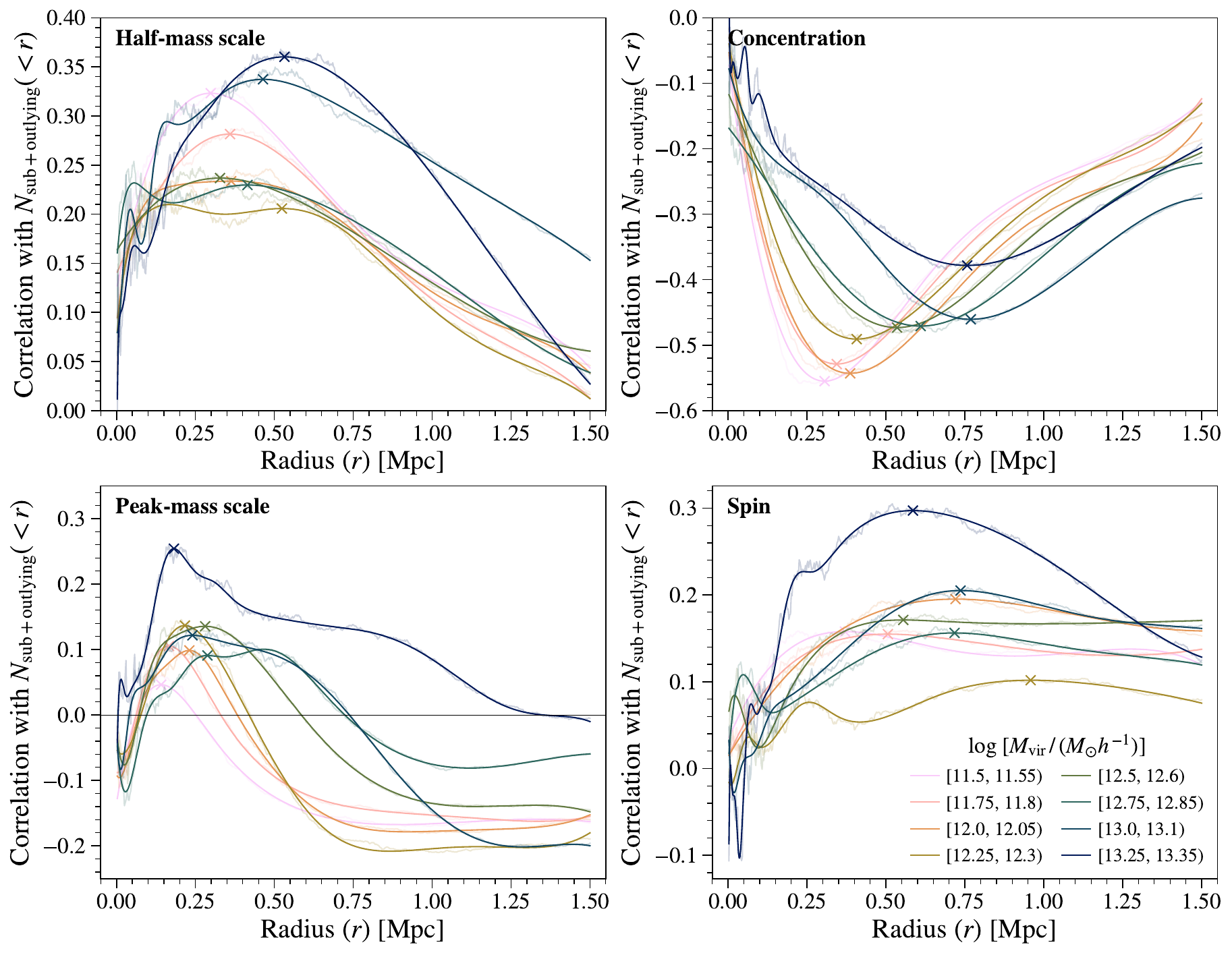}
\caption{Spearman correlation between each of the four host halo properties (half-mass scale, \textit{upper left}; concentration, \textit{upper right}; peak-mass scale, \textit{bottom left}; spin, \textit{bottom right}) and the cumulative number of subhalos and outlying halos within radius $r$, similar to the left panel of \autoref{fig:correlation}. The subhalos and outlying halos are always selected with a \vmaxmpeak{} threshold of 20\,\kms. 
The different colors (shades) represent different host halo mass ranges. The semi-transparent and solid curves represent the unsmoothed and smoothed correlation--radius relations. The cross marks represent the location at which the correlation is strongest in the smoothed relation.
\label{fig:correlation-masses}
}
\end{figure*}

\begin{figure*}[tb!]
\centering
\includegraphics[width=\linewidth]{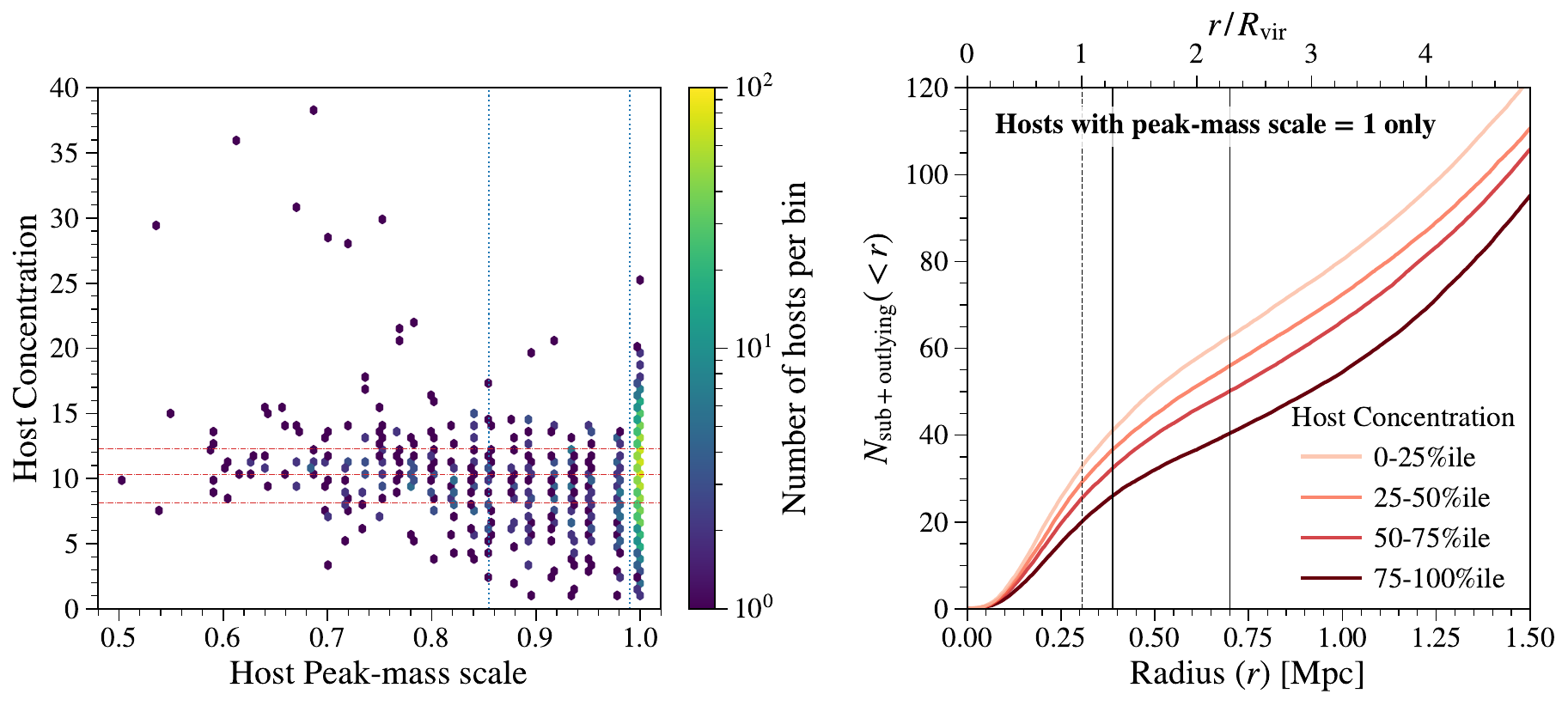}
\caption{\textit{Left:} Joint distribution of the host halo concentration and peak-mass scale in hexagonal bins. The color indicates the number of hosts per bin. Blue dotted vertical lines divide the three peak-mass scale ranges as shown in the lower left panel of \autoref{fig:radial-dist}. Red dashed horizontal lines divide the concentration quartiles. 
\textit{Right:} Same as the upper right panel of \autoref{fig:radial-dist}, but with only host halos whose peak-mass scale is 1 (i.e., $z=0$). The vertical lines are shown at the same locations as in the upper right panel of \autoref{fig:radial-dist} for comparison. 
\label{fig:c_vs_ampeak}
}
\end{figure*}

Here we present a few additional figures for interested readers. 

\autoref{fig:radial-dist-log} is identical to \autoref{fig:radial-dist}, except that the $y$-axis is shown in the logarithmic scale. In this view, parallel lines indicate a multiplicative difference between them. For example, for the case of host spin, the difference in the numbers of outlying halos around 0.7 to 0.9~Mpc appears to be more multiplicative than additive.

\autoref{fig:correlation-masses} shows the correlation between a halo property and the number of subhalos within a radius $r$, similar to the left panel of \autoref{fig:correlation}.  However, \autoref{fig:correlation} only shows this result for a specific host halo mass ($10^{12-12.05}$\,\msun\,\perh). In \autoref{fig:correlation-masses}, we show the correlation for eight host halo mass bins: $\log [ \mvir / (\msun\,\perh) ] \in [11.5, 11.55)$,
$[11.75, 11.8)$,
$[12, 12.05)$,
$[12.25, 12.35)$,
$[12.5, 12.6)$,
$[12.75, 12.85)$,
$[13, 13.1)$, and 
$[13.25, 13.35)$. 
Each curve shows how the correlation varies with radius for one specific host halo mass. 
The four panels are for the four different halo properties we have analyzed. 

In \autoref{fig:correlation-masses} we also identify the radius at which the correlation is the strongest. These strongest-correlation radii are marked with crosses. Because the correlation--radius curves are noisy (due to the Poisson noise in the counting statistics), we smooth the correlation--radius curves before identifying the strongest-correlation radii. Both the smoothed and unsmoothed curves are shown in \autoref{fig:correlation-masses}.
From the crosses marked in \autoref{fig:correlation-masses}, we directly observe how the strongest-correlation radii evolve with halo mass for each halo property we analyzed. These evolution trends are summarized in \autoref{fig:mass-trend} and discussed in detail in \autoref{sec:radius-trend}. 

The left panel of \autoref{fig:c_vs_ampeak} shows the joint distribution of the host halo concentration and peak-mass scale, for the host halo sample in the Milky Way-mass bin ($\mvir = 10^{12-12.05}\,\msun\,\perh$). Because we apply an isolation criterion (no bigger halo within 1.5\,Mpc) on our host halo sample, most of the host halos have their peak-mass scale at $a=1$ (i.e., they reach their peak mass at present day). The main feature of interest in this figure is that the most high-concentration halos (with a concentration parameter greater than 20) have early peak-mass scales. Halos with early peak-mass scales have experienced some mass stripping in the past, even though they may be isolated today. 

This feature explains the result we saw in \autoref{fig:radial-dist}, where the halos with early peak-mass scales or high concentration have a distinct subhalo radial profile (low in the inner region but high in the outer region). As we discussed in \autoref{sec:assembly-bias}, this distinct subhalo radial profile is a result of the combined effect of occupation variation and halo assembly bias. Host halos that have experienced stripping in the past are a major contributor to the halo assembly bias effect \citep{2020MNRAS.493.4763M}, resulting in a high number of subhalos in the outer region. And since these host halos also have a higher concentration (as shown in \autoref{fig:c_vs_ampeak}), they tend to have fewer subhalos in the inner region. 

In fact, when we remove hosts with peak-mass scales less than $a=1$ and remake the radial profiles, which is now shown in the right panel of \autoref{fig:c_vs_ampeak}, we can see that the outlier behavior from the host halos with the highest concentration disappears. The lines from quartiles now remain mostly parallel in the outer region.
In other words, the correlation strength as a function of radius (that we investigated in \autoref{sec:correlation}) will roughly stay constant after reaching its peak value. 
However, we note that it is nearly impossible to exclude hosts with peak-mass scales less than $a=1$ from an observational perspective. 

\end{document}